\documentclass[twocolumn,showpacs,%
  nofootinbib,aps,superscriptaddress,%
  eqsecnum,prd,notitlepage,showkeys,10pt]{revtex4-1}

\usepackage{amssymb}
\usepackage{amsmath}
\usepackage{graphicx}
\usepackage{dcolumn}
\usepackage{hyperref}
\usepackage{multirow}

\begin{document}

\title{Accurate absolute core-electron binding energies of molecules, solids and surfaces from first-principles calculations}
\author{J. Matthias Kahk}
\affiliation{Department of Materials, Imperial College London, South Kensington, London SW7 2AZ, United Kingdom}
\author{Johannes Lischner}
\affiliation{Department of Physics and Department of Materials, and the Thomas Young Centre for Theory and Simulation of Materials, Imperial College London, London SW7 2AZ, United Kingdom}
\email{j.lischner@imperial.ac.uk}

\begin{abstract}
Core-electron x-ray photoelectron spectroscopy is a powerful technique for studying the electronic structure and chemical composition of molecules, solids and surfaces. However, the interpretation of measured spectra and the assignment of peaks to atoms in specific chemical environments is often challenging. Here, we address this problem and introduce a parameter-free computational approach for calculating absolute core-electron binding energies. In particular, we demonstrate that accurate absolute binding energies can be obtained from the total energy difference of the ground state and a state with an explicit core hole when exchange and correlation effects are described by a recently developed meta-generalized gradient approximation and relativistic effects are included even for light elements. We carry out calculations for molecules, solids and surface species and find excellent agreement with available experimental measurements. For example, we find a mean absolute error of only 0.16 eV for a reference set of 103 molecular core-electron binding energies. The capability to calculate accurate absolute core-electron binding energies will enable new insights into a wide range of chemical surface processes that are studied by x-ray photoelectron spectroscopy.

\end{abstract}

\maketitle

\section{Introduction}

Core-level x-ray photoelectron spectroscopy (XPS) measures the energies required to remove core electrons from atoms in a given sample. As these energies depend sensitively on the atom's chemical environment, XPS is a powerful method for chemical analysis. In particular, core-level XPS finds widespread use in the characterization of the surfaces of solids, and insights gained from XPS measurements play a crucial role in developing our understanding of various surface chemical processes, including heterogeneous catalyis \cite{murugappan_operando_2018,goulas_fundamentals_2019,kattel_active_2017,feng_understanding_2013}, the formation of electrified interfaces \cite{favaro_unravelling_2016,booth_offset_2017,axnanda_using_2015}, corrosion and degradation \cite{bouanis_corrosion_2016,zarrok_corrosion_2012,parkinson_iron_2016,orlando_environmental_2016}, and adhesion \cite{yoshida_comparative_2004,noeske_plasma_2004,wang_verification_2013}. A key challenge in applying XPS to complex materials is that it is often difficult to assign peaks in the XPS spectrum to specific chemical environments. Overcoming this peak assignment problem is critical in order to maximize the chemical insight gained from experimental XPS measurements.

Theoretical calculations of core-electron binding energies of atoms in different chemical environments have the potential to guide the interpretation of XPS spectra and several approaches have been developed to achieve this goal. The most common approaches are the $\Delta$SCF method where the core-electron binding energy is calculated as the total energy difference between the ground state and the final state with a core hole,\cite{bagus_self-consistent-field_1965} and the related Slater-Janak transition state method \cite{slater_statistical_1972,williams_generalization_1975}. These techniques yield \emph{relative} core-electron binding energies, or binding energy shifts, that are in good agreement with experimental measurements for free molecules \cite{vines_prediction_2018,takahata_accurate_2010-1,tolbatov_comparative_2014,takahata_estimation_2005,pueyo_bellafont_performance_2016-1,pueyo_bellafont_predicting_2017}. While calculated binding energy shifts are often very useful for the interpretation of XPS spectra, their use requires the existence of well-defined core-level reference energies which is not always guaranteed. It is therefore highly desirable to also calculate \emph{absolute} core-electron binding energies, but these are often found to differ by multiple electron volts from measured values. Some works have reported the prediction of accurate absolute core-electron binding energies ,\cite{takahata_accurate_2010,oakley_dft/mix:_2018,tolbatov_comparative_2014,cavigliasso_accurate_1999} but this typically relies on the fortuitous cancellation of errors arising from incomplete basis sets, limitations in the treatment of exchange and correlation effects and the neglect or empirical treatment of relativistic effects \cite{vines_prediction_2018}. The reliance on error cancellations ultimately limits the generality and the accuracy of these methods.

A variant of the $\Delta$SCF scheme that overcomes some of the limitations of earlier studies was recently proposed by Pueyo Bellafont and coworkers \cite{pueyo_bellafont_performance_2016}. Using the TPSS meta-GGA exchange-correlation functional and fully uncontracted gaussian basis sets, absolute 1s core-electron binding energies of B, C, N, O and F atoms in free molecules were obtained that agree with experimental gas phase measurements to within a mean absolute error of 0.21 eV, provided that corrections due to relativistic effects (ranging from 0.06 eV for B 1s to 0.75 eV for F 1s) are added to the calculated values. 

As most XPS experiments are carried out on solids and surfaces, methods to calculate core-electron binding energies in such systems must be developed and several new approaches have been proposed in recent years. For example, Ozaki et al. introduced a formalism based on periodic density-functional theory (DFT) calculations where the effect of the core hole is simulated by introducing a penalty functional and the spurious interaction between periodically repeated core holes is removed by the exact Coulomb cutoff method \cite{ozaki_absolute_2017}. However, this method employs the frozen core approximation and thus neglects the screening effect from the relaxations of other core electrons resulting in significant quantitative inaccuracies. For example, for some simple molecules, such as NH$_3$ and N$_2$H$_4$, the calculated values differ by -0.9 eV and -1.28 eV from experimental data, respectively. Another method for calculating absolute core-electron binding energies in extended systems is based on the GW approach \cite{aoki_accurate_2018}. However, the high computational cost of GW calculations makes its application to chemically complex systems challenging. Moreover, preliminary tests of the GW method have failed to match the accuracy reported for $\Delta$SCF calculations using the TPSS functional \cite{van_setten_assessing_2018,golze_core-level_2018,pueyo_bellafont_performance_2016}.

In this work, we propose a novel method based on the all-electron $\Delta$SCF for calculating absolute core-electron binding energies of molecules, surfaces and solids. In particular, we calculate total energies of the ground state and the final state which contains an explicit, spin-polarized core hole using DFT with the Strongly Constrained and Appropriately Normed (SCAN) exchange-correlation functional \cite{sun_strongly_2015}. The SCAN functional is a non-empirical semi-local meta-GGA functional that obeys several exact constraints and combines generality, affordability and good performance in benchmark calculations \cite{sun_strongly_2015,sun_accurate_2016,isaacs_performance_2018,marianski_assessing_2016,goerigk_look_2017,tozer_molecular_2018}. Relativistic effects are included self-consistently via the scaled Zeroth Order Regular Approximation (ZORA).\cite{van_lenthe_relativistic_1994} For orbital eigenvalues in Hartree-Fock calculations of free atoms, the scaled ZORA method yields accurate results (error relative to Dirac-Fock calculations less than 0.1 eV) for valence and shallow core levels of all elements and for all core and valence levels of light elements.\cite{van_lenthe_relativistic_1994,faas_zora_1995,dyall_relativistic_1999,klopper_improved_2000}. Extended systems are described using finite cluster models containing several hundred atoms, thereby allowing the $\Delta$SCF method to be directly applied to solids and surface species.

\section{Results and discussion}

\begin{figure*}
	\centering
	\includegraphics[width=7.00in]{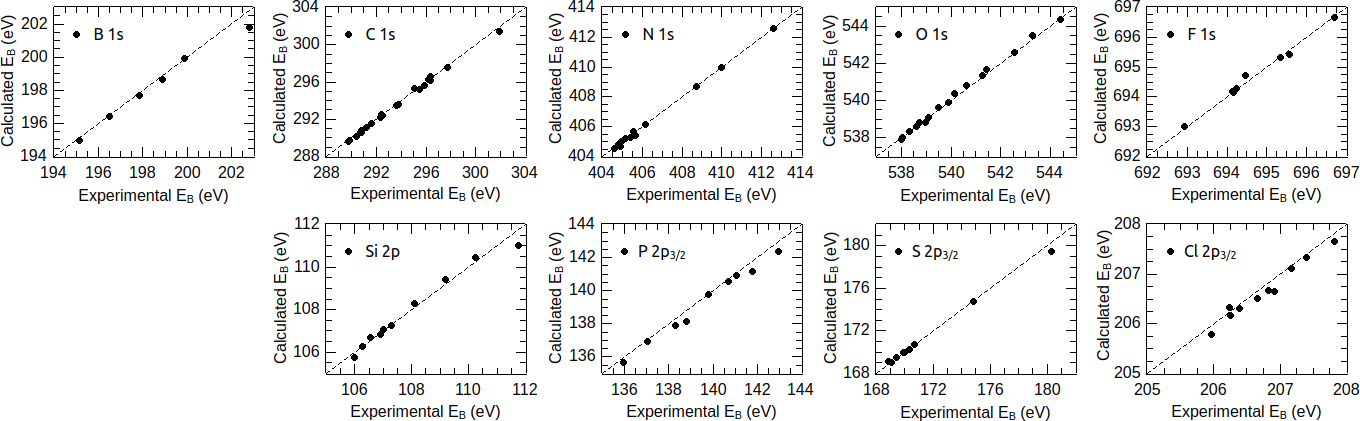}
	\caption{A comparison of experimental and calculated core-electron binding energies for free molecules.}
	\label{Fig_gas_phase}
\end{figure*}

To assess the accuracy of our approach for free molecules, we carry out calculations for 75 molecules with a total of 103 core-electron binding energies, see Supplementary Information for a full list of all molecules. This data set contains 1s binding energies for the elements B, C, N, O and F and 2p binding energies for the elements Si, P, S and Cl. To minimize the experimental error, we only included molecules whose core-electron binding energies were measured at least twice with the reported binding energies differing by no more than 0.3 eV. Subject to these criteria, the arithmetic averages of all reported experimental binding energies were then chosen as the reference values. In addition, the “weighted average” binding energies from Cavigliasso have also been included in the reference set \cite{cavigliasso_accurate_1999}. The molecular structures were relaxed before the $\Delta$SCF calculations of the core-electron binding energies. 
Figure~\ref{Fig_gas_phase} shows the calculated core-electron binding energies for free molecules and compares them to experimental gas phase measurements, see also Table~\ref{Table_molecules}. Our approach yields excellent agreement with experimental measurements with a mean error of only -0.09 eV and a mean absolute error of 0.16 eV. For 95 out of the 103 core-level binding energies of the data set, the agreement with experiment is better than 0.3 eV. Given that core-electron binding energies for different chemical environments range over several eV for each of the considered 1s and 2p core levels, an accuracy better than 0.3 eV for the vast majority of cases means that the calculated binding energies are a reliable guide for the interpretation of experimental spectra. For example, with very few exceptions, the energy ordering of almost all of the experimental datapoints is predicted correctly. To compare the performance of our approach to the method of Ref. [\citenum{ozaki_absolute_2017}], we have carried out calculations for the molecular dataset of Ref. [\citenum{ozaki_absolute_2017}]. We find that our approach yields a significantly smaller mean absolute error (0.19 eV vs 0.52 eV) and also a significantly smaller maximum error (0.70 eV vs 1.28 eV).

\begin{table*}
	\begin{tabular}{ c c c c c c c c c c c}
		\hline 
		& B 1s & C 1s & N 1s & O 1s & F 1s & Si 2p & P 2p$_{3/2}$ & S 2p$_{3/2}$ & Cl 2p$_{3/2}$ & Total \\ 
		\hline 
		Datapoints & 6 & 22 & 13 & 16 & 8 & 10 & 9 & 9 & 10 & 103 \\
		ME (eV) & -0.29 & -0.12 & -0.02 & 0.00 & 0.02 & -0.03 & -0.33 & -0.05 & -0.14 & -0.09 \\
		MAE (eV) & 0.37 & 0.16 & 0.08 & 0.11 & 0.09 & 0.19 & 0.33 & 0.18 & 0.15 & 0.16 \\
		\hline 
	\end{tabular} 	
	\caption{Summary of the calculated core-electron binding energies for free molecules and comparison to experiment. ME = mean error (theory - experiment). MAE = mean absolute error.}
	\label{Table_molecules}
\end{table*}

In the free molecule calculations, the eight core levels that differ most from the measured values are the B 1s core level in BF$_{3}$ (theoretical binding energy - experimental binding energy = -1.04 eV), S 2p$_{3/2}$ in SF$_{6}$ (-0.84 eV), Si 2p in SiF$_{4}$ (-0.75 eV), P 2p$_{3/2}$ in P(CF$_{3}$)$_{3}$ (-0.68 eV), P 2p$_{3/2}$ in PF$_{3}$ (-0.63 eV), P 2p$_{3/2}$ in POF$_{3}$ (-0.61 eV), C 1s in CF$_{4}$ (-0.45 eV) and P 2p$_{3/2}$ in P(OCH$_{3}$)$_{3}$ (-0.37 eV). The relevant atoms in these molecules are in high oxidation states and we speculate that the observed errors are a consequence of the well-known difficulties of (semi-)local exchange-correlation functionals in describing systems with large charge transfer \cite{perdew_density-functional_1982,cohen_insights_2008}.

To calculate absolute core-electron binding energies of extended systems, we follow a two-step procedure that we have used to predict binding energy shifts in a previous study.\cite{kahk_core_2018} First, the atomic positions are relaxed using a periodic model of the (2D or 3D) material. Next, a cluster is cut from the periodic structure and the $\Delta$SCF method is used to calculate core-electron binding energies. In contrast to gas-phase experiments where the measured binding energies are referenced to the vacuum level (corresponding to a final state with one electron removed from the sample and promoted to the vacuum level), the binding energies obtained in XPS measurements on solids and surfaces are referenced to the Fermi level of the material. The corresponding core-electron binding energies can be obtained directly from a $\Delta$SCF calculation where the core electron is promoted to an occupied state at the Fermi level (instead of being removed from the sample). In other words, the core-electron binding energy is calculated as the total energy difference between the ground state and a neutral final state with a core hole and an extra electron in the conduction band. This approach is valid as long as there are empty electronic states at the Fermi level, which is true for the metallic systems considered in this work. For systems with a band gap, the valence band maximum is a better reference point than the Fermi level. In that case, the core-electron binding energy can be calculated as the difference between the absolute core-electron binding energy referenced to the vacuum level, and the energy required to remove an electron from the valence band maximum which can be obtained from a separate $\Delta$SCF calculation.

\begin{figure}
	\centering
	\includegraphics[width=1.8in]{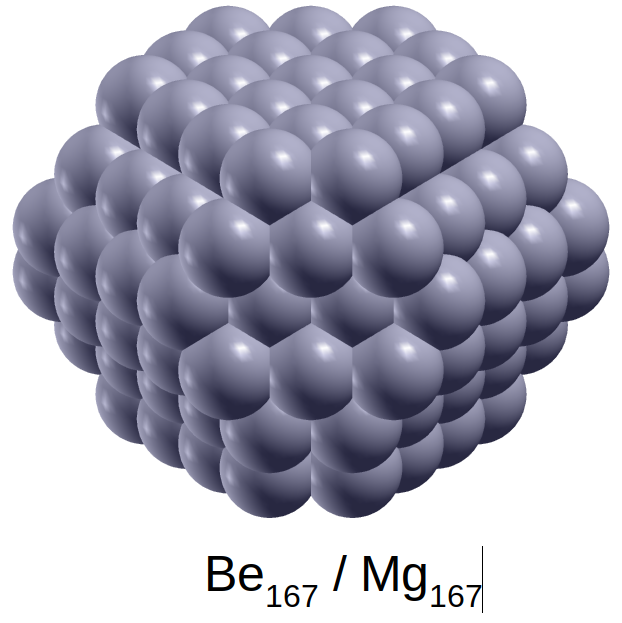}
	\caption{The cluster of 167 atoms used to calculate the 1s core-electron binding energy in Be and Mg. The core hole is localized on an atom at the centre of the cluster.}
	\label{Fig_Be_cluster}
\end{figure}

We have used the approach outlined above to calculate absolute core-electron binding energies in two elemental metals: Mg and Be. Specifically, we first carried out periodic DFT calculations using the SCAN functional to determine the equilibrium bulk lattice constants of Be and Mg, see Supplementary Materials. Next, finite clusters containing 167 atoms were constructed from the bulk structure (see Fig.~\ref{Fig_Be_cluster}) and the 1s core-electron binding energies were calculated using the $\Delta$SCF approach with the core hole being localized at an atom in the center of the cluster. Our results are summarized in Table \ref{Table_bulk_metals}. For both Be and Mg, the calculated core-electron binding energies are within 0.3 eV of experimental values. \cite{powell_recommended_2012,powell_elemental_1995,ley_many-body_1975,fuggle_xps_1977,jennison_calculation_1984,darrah_thomas_valence_1986,peng_reactions_1988,yoshimura_degradation_2007,fischer_hydrogen_1991}

\begin{table}
	\begin{tabular}{ c c c }
		\hline 
		& Be & Mg \\
		\hline
		Theoretical $E_B$ (eV) & 111.57 & 1303.07 \\
		Experimental $E_B$ (eV) & 111.82 & 1303.2 \\
		\hline 
	\end{tabular} 	
	\caption{Calculated and experimental 1s binding energies ($E_B$) in metallic beryllium and magnesium}
	\label{Table_bulk_metals}
\end{table}

\begin{figure*}
	\centering
	\includegraphics[width=5.00in]{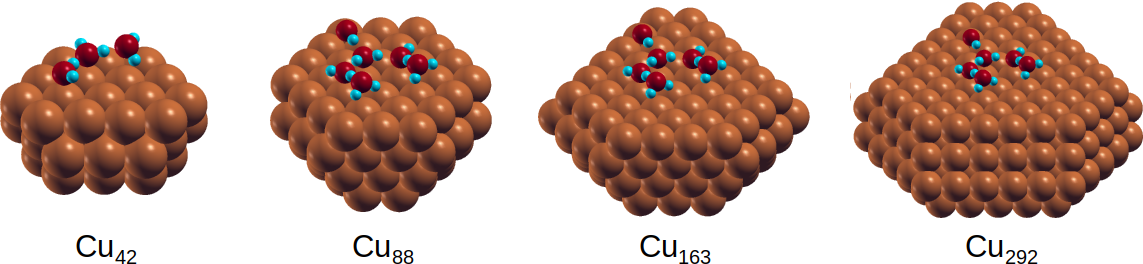}
	\caption{The four Cu clusters that were used in $\Delta$SCF calculations of the O 1s core-electron binding energy of adsorbed H$_{2}$O.}
	\label{Fig_Cu_clusters}
\end{figure*}

Finally we turn our attention to the prediction of core-electron binding energies of adsorbed molecules. We have carried out calculations for four molecules, H$_2$O, OH, CO and HCOO, on Cu(111). In these calculations, a Cu cluster containing 163 Cu atoms, shown in Figure \ref{Fig_Cu_clusters}, was used to model the Cu(111) surface. The relaxed geometries of the adsorbates on the surface were taken from our previous study \cite{kahk_core_2018}. For CO on Cu(111), the ``top" adsorption site was used (where the CO sits directly above one of the Cu atoms). For water, an adsorbed H$_2$O molecule hydrogen bonded to two other adsorbed water molecules was considered. Table~\ref{Table_adsorbates} compares the calculated absolute core-electron binding energies for the molecules adsorbed to the Cu(111) surface with experimental measurements. For these adsorbed molecules, ``reference quality" experimental data is in general not available, making it difficult to judge the accuracy of the calculated values in a quantitative manner. However, in so far as can be determined, the agreement between theory and experiment is good. The calculated O 1s binding energies in adsorbed H$_2$O, OH and HCOO are all within 0.35 eV of the experimental values; the calculated C 1s binding energies of adsorbed CO and adsorbed HCOO are within 0.5 eV of the closest experimental data points, and the calculated O 1s binding energy in adsorbed CO lies in between two experimental values that differ from each other by almost 2 eV.

To establish that the obtained results are converged with respect to cluster size, we have calculated the O 1s binding energy of adsorbed H$_2$O on four different Cu clusters consisting of 42, 88, 163 and 292 Cu atoms, respectively. These clusters are shown in Figure \ref{Fig_Cu_clusters}. Table~\ref{Table_cluster_size} shows that the calculated core electron binding energies for the different cluster sizes vary by less than 0.15 eV.

\begin{table}
	\begin{tabular}{ c c c c}
		\hline 
		Core level & Theor. $E_{B}$ & Exp. $E_{B}$ & \\
		& (eV) & (eV) & \\
		\hline
		\multirow{2}{*}{H$_{2}$O O 1s} & \multirow{2}{*}{532.95} & 533.0 & [\!\!\!\citenum{roberts_low_2014}] \\
		& & 532.4 & [\!\!\!\citenum{favaro_subsurface_2017}] \vspace{1mm} \\
		
		OH O 1s & 531.15 & 531.50 & [\!\!\!\citenum{mudiyanselage_importance_2013}] \vspace{1mm} \\
		\multirow{2}{*}{CO C 1s} & \multirow{2}{*}{285.70} & 286.1 & [\!\!\!\citenum{eren_catalyst_2015}] \\
		& & 286.2 & [\!\!\!\citenum{mudiyanselage_importance_2013}] \vspace{1mm}\\
		\multirow{2}{*}{CO O 1s} & \multirow{2}{*}{532.52} & 531.5 & [\!\!\!\citenum{eren_catalyst_2015}]\\
		& & 533.4 & [\!\!\!\citenum{mudiyanselage_importance_2013}] \vspace{1mm}\\
		\multirow{3}{*}{HCOO C 1s} & \multirow{3}{*}{286.82} & 287.3 & [\!\!\!\citenum{favaro_subsurface_2017}] \\
		& & 288.2 & [\!\!\!\citenum{nakamura_x-ray_1997}] \\
		& & 289.75 & [\!\!\!\citenum{yang_fundamental_2010}] \vspace{1mm}\\
		HCOO O 1s & 531.25 & 531.5 & [\!\!\!\citenum{nakamura_x-ray_1997}] \\
		\hline 
	\end{tabular} 	
	\caption{A comparison of experimental and calculated core-electron binding energies for adsorbates on Cu(111).}
	\label{Table_adsorbates}
\end{table}

\begin{table}
	\begin{tabular}{ c c c }
		\hline 
		\multirow{2}{*}{Species} & Calculated \\
		& O 1s $E_{B}$ (eV) \\
		\hline
		H$_{2}$O on Cu$_{42}$ & 532.97 \\
		H$_{2}$O on Cu$_{88}$ & 532.85 \\
		H$_{2}$O on Cu$_{163}$ & 532.95 \\
		H$_{2}$O on Cu$_{292}$ & 532.84 \\
		\hline 
	\end{tabular} 	
	\caption{Dependence of the calculated absolute core-electron binding energy of an adsorbed water molecule on the size of the Cu cluster that is used to simulate the Cu(111) surface.}
	\label{Table_cluster_size}
\end{table}

\section{Conclusions}

We find that the $\Delta$SCF approach yields accurate \emph{absolute} core-electron binding energies for molecules, solids and surfaces when the SCAN exchange-correlation energy functional is employed in conjunction with the scaled ZORA treatment of relativistic effects. Specifically, we find that our calculated binding energies agree with experiments to within 0.3 eV. This accuracy is usually sufficient to guide the interpretation of experimental XPS spectra and overcome the peak assignment problem that often limits the amount of information that can be extracted from XPS studies of complex materials. A shortcoming of the present approach is the perturbative treatment of spin-orbit coupling which is applied after the self-consistent field calculation. This reduces the accuracy of the approach for heavier elements. Future work will include a fully self-consistent treatment of spin-orbit coupling that will allow the accurate description of all elements in the periodic table.

\section{Methods}

All of the calculations reported in this work were carried out using the FHI-aims electronic structure code in which the Kohn-Sham wavefunctions are constructed as linear combinations of numerical atom-centred orbitals.\cite{blum_ab_2009,havu_efficient_2009,strange_automatic_2001} The geometries of the free molecules and bulk Be and Mg were relaxed until the forces on all atoms were less than 5.0 $\cdot~10^{-3}$ eV/$\textnormal{\AA}$. Variable-cell optimization was used for Be and Mg. The FHI-aims default ``tight" basis sets were used in the geometry optimizations.\cite{blum_ab_2009} For metallic Be and Mg, the Brillouin zone was sampled using a 12 x 12 x 8 grid. For the adsorbates on Cu(111), relaxed geometries from our previous study were used.\cite{kahk_core_2018} In the $\Delta$SCF calculations, special basis sets were used for the atoms where a core hole was created. These basis sets were constructed by adding additional, tighter core wavefunctions to the default basis sets of FHI-aims in order to permit the relaxation of other core electrons in the presence of a core hole. Tests on simple molecules showed that the calculated core-electron binding energies obtained using the numerical basis sets were within 0.08 eV of the values obtained for the same systems using large, uncontracted gaussian basis sets derived from the pcJ-3 basis sets of Jensen.\cite{jensen_optimum_2010} Full details of the basis sets used in this work are given in the supplementary information. When calculating the total energy of the final state, the core hole was introduced by constraining the occupancy of a specific Kohn-Sham state in one of the spin channels. In cases where the molecule contains a number of atoms of the same element, the localization of the core hole at a specific atom was ensured by introducing a fictitious extra charge of 0.1 $e$ during the first step of the self-consistent field cycle at the desired site which attracts the core hole. The fictitious charge was removed immediately afterwards and the constrained self-consistent field calculations were run in the usual manner. 

All of the reported calculations were scalar relativistic, with optional post-self-consistent perturbative spin-orbit coupling.\cite{huhn_one-hundred-three_2017} Therefore, the calculated 2p core-electron binding energies correspond to the weighted average of the 2p$_{1/2}$ and 2p$_{3/2}$ states. For comparisons to experimental data, the as-obtained 2p binding energies were used for Si. For P, S, and Cl, the position of the 2p$_{3/2}$ peak is usually reported experimentally. The theoretical 2p$_{3/2}$ binding energies were obtained by subtracting 1/3 of the experimental spin-orbit splitting (0.29 eV for P 2p, 0.387 eV for S 2p, and 0.533 eV for Cl 2p) from the $\Delta$SCF value. Very similar binding energies (to within less than 0.05 eV) are obtained if theoretical spin-orbit splittings, as determined from the eigenvalue difference after the perturbative SOC calculations, are used instead.

\section{Acknowledgement}

J.M.K. and J.L. acknowledge support from EPRSC under Grant No. EP/R002010/1. Via J.L.'s membership of the UK's HEC Materials Chemistry Consortium, which is funded by EPSRC (EP/L000202), this work used the ARCHER UK National Supercomputing Service.


\begin{thebibliography}{10}
	
	\bibitem{murugappan_operando_2018}
	K.~Murugappan, E.~M. Anderson, D.~Teschner, T.~E. Jones, K.~Skorupska, and
	Y.~Román-Leshkov, ``Operando {NAP}-{XPS} unveils differences in {MoO}3 and
	{Mo}2c during hydrodeoxygenation,'' {\em Nature Catalysis}, vol.~1,
	pp.~960--967, Dec. 2018.
	
	\bibitem{goulas_fundamentals_2019}
	K.~A. Goulas, A.~V. Mironenko, G.~R. Jenness, T.~Mazal, and D.~G. Vlachos,
	``Fundamentals of {C}–{O} bond activation on metal oxide catalysts,'' {\em
		Nature Catalysis}, vol.~2, pp.~269--276, Mar. 2019.
	
	\bibitem{kattel_active_2017}
	S.~Kattel, P.~J. Ramírez, J.~G. Chen, J.~A. Rodriguez, and P.~Liu, ``Active
	sites for {CO} $_{\textrm{2}}$ hydrogenation to methanol on {Cu}/{ZnO}
	catalysts,'' {\em Science}, vol.~355, pp.~1296--1299, Mar. 2017.
	
	\bibitem{feng_understanding_2013}
	N.~Feng, Q.~Wang, A.~Zheng, Z.~Zhang, J.~Fan, S.-B. Liu, J.-P. Amoureux, and
	F.~Deng, ``Understanding the {High} {Photocatalytic} {Activity} of ({B},
	{Ag})-{Codoped} {TiO} $_{\textrm{2}}$ under {Solar}-{Light} {Irradiation}
	with {XPS}, {Solid}-{State} {NMR}, and {DFT} {Calculations},'' {\em Journal
		of the American Chemical Society}, vol.~135, pp.~1607--1616, Jan. 2013.
	
	\bibitem{favaro_unravelling_2016}
	M.~Favaro, B.~Jeong, P.~N. Ross, J.~Yano, Z.~Hussain, Z.~Liu, and E.~J.
	Crumlin, ``Unravelling the electrochemical double layer by direct probing of
	the solid/liquid interface,'' {\em Nature Communications}, vol.~7, Dec. 2016.
	
	\bibitem{booth_offset_2017}
	S.~G. Booth, A.~M. Tripathi, I.~Strashnov, R.~A.~W. Dryfe, and A.~S. Walton,
	``The offset droplet: a new methodology for studying the solid/water
	interface using x-ray photoelectron spectroscopy,'' {\em Journal of Physics:
		Condensed Matter}, vol.~29, p.~454001, Nov. 2017.
	
	\bibitem{axnanda_using_2015}
	S.~Axnanda, E.~J. Crumlin, B.~Mao, S.~Rani, R.~Chang, P.~G. Karlsson, M.~O.~M.
	Edwards, M.~Lundqvist, R.~Moberg, P.~Ross, Z.~Hussain, and Z.~Liu, ``Using
	“{Tender}” {X}-ray {Ambient} {Pressure} {X}-{Ray} {Photoelectron}
	{Spectroscopy} as {A} {Direct} {Probe} of {Solid}-{Liquid} {Interface},''
	{\em Scientific Reports}, vol.~5, Sept. 2015.
	
	\bibitem{bouanis_corrosion_2016}
	M.~Bouanis, M.~Tourabi, A.~Nyassi, A.~Zarrouk, C.~Jama, and F.~Bentiss,
	``Corrosion inhibition performance of
	2,5-bis(4-dimethylaminophenyl)-1,3,4-oxadiazole for carbon steel in {HCl}
	solution: {Gravimetric}, electrochemical and {XPS} studies,'' {\em Applied
		Surface Science}, vol.~389, pp.~952--966, Dec. 2016.
	
	\bibitem{zarrok_corrosion_2012}
	H.~Zarrok, A.~Zarrouk, B.~Hammouti, R.~Salghi, C.~Jama, and F.~Bentiss,
	``Corrosion control of carbon steel in phosphoric acid by purpald –
	{Weight} loss, electrochemical and {XPS} studies,'' {\em Corrosion Science},
	vol.~64, pp.~243--252, Nov. 2012.
	
	\bibitem{parkinson_iron_2016}
	G.~S. Parkinson, ``Iron oxide surfaces,'' {\em Surface Science Reports},
	vol.~71, pp.~272--365, Mar. 2016.
	
	\bibitem{orlando_environmental_2016}
	F.~Orlando, A.~Waldner, T.~Bartels-Rausch, M.~Birrer, S.~Kato, M.-T. Lee,
	C.~Proff, T.~Huthwelker, A.~Kleibert, J.~van Bokhoven, and M.~Ammann, ``The
	{Environmental} {Photochemistry} of {Oxide} {Surfaces} and the {Nature} of
	{Frozen} {Salt} {Solutions}: {A} {New} in {Situ} {XPS} {Approach},'' {\em
		Topics in Catalysis}, vol.~59, pp.~591--604, Mar. 2016.
	
	\bibitem{yoshida_comparative_2004}
	Y.~Yoshida, K.~Nagakane, R.~Fukuda, Y.~Nakayama, M.~Okazaki, H.~Shintani,
	S.~Inoue, Y.~Tagawa, K.~Suzuki, J.~De~Munck, and B.~Van~Meerbeek,
	``Comparative {Study} on {Adhesive} {Performance} of {Functional}
	{Monomers},'' {\em Journal of Dental Research}, vol.~83, pp.~454--458, June
	2004.
	
	\bibitem{noeske_plasma_2004}
	M.~Noeske, J.~Degenhardt, S.~Strudthoff, and U.~Lommatzsch, ``Plasma jet
	treatment of five polymers at atmospheric pressure: surface modifications and
	the relevance for adhesion,'' {\em International Journal of Adhesion and
		Adhesives}, vol.~24, pp.~171--177, Apr. 2004.
	
	\bibitem{wang_verification_2013}
	Y.~Wang, J.~Xue, Q.~Wang, Q.~Chen, and J.~Ding, ``Verification of
	{Icephobic}/{Anti}-icing {Properties} of a {Superhydrophobic} {Surface},''
	{\em ACS Applied Materials \& Interfaces}, vol.~5, pp.~3370--3381, Apr. 2013.
	
	\bibitem{bagus_self-consistent-field_1965}
	P.~S. Bagus, ``Self-{Consistent}-{Field} {Wave} {Functions} for {Hole} {States}
	of {Some} {Ne}-{Like} and {Ar}-{Like} {Ions},'' {\em Physical Review},
	vol.~139, pp.~A619--A634, Aug. 1965.
	
	\bibitem{slater_statistical_1972}
	J.~C. Slater, ``Statistical {Exchange}-{Correlation} in the {Self}-{Consistent}
	{Field},'' in {\em Advances in {Quantum} {Chemistry}}, vol.~6, pp.~1--92,
	Elsevier, 1972.
	
	\bibitem{williams_generalization_1975}
	A.~R. Williams, R.~A. deGroot, and C.~B. Sommers, ``Generalization of
	{Slater}’s transition state concept,'' {\em The Journal of Chemical
		Physics}, vol.~63, pp.~628--631, July 1975.
	
	\bibitem{vines_prediction_2018}
	F.~Viñes, C.~Sousa, and F.~Illas, ``On the prediction of core level binding
	energies in molecules, surfaces and solids,'' {\em Physical Chemistry
		Chemical Physics}, vol.~20, no.~13, pp.~8403--8410, 2018.
	
	\bibitem{takahata_accurate_2010-1}
	Y.~Takahata and A.~D.~S. Marques, ``Accurate core-electron binding energy
	shifts from density functional theory,'' {\em Journal of Electron
		Spectroscopy and Related Phenomena}, vol.~178-179, pp.~80--87, May 2010.
	
	\bibitem{tolbatov_comparative_2014}
	I.~Tolbatov and D.~M. Chipman, ``Comparative study of {Gaussian} basis sets for
	calculation of core electron binding energies in first-row hydrides and
	glycine,'' {\em Theoretical Chemistry Accounts}, vol.~133, Oct. 2014.
	
	\bibitem{takahata_estimation_2005}
	Y.~Takahata and D.~P. Chong, ``Estimation of {Hammett} sigma constants of
	substituted benzenes through accurate density-functional calculation of
	core-electron binding energy shifts,'' {\em International Journal of Quantum
		Chemistry}, vol.~103, no.~5, pp.~509--515, 2005.
	
	\bibitem{pueyo_bellafont_performance_2016-1}
	N.~Pueyo~Bellafont, G.~Álvarez Saiz, F.~Viñes, and F.~Illas, ``Performance of
	{Minnesota} functionals on predicting core-level binding energies of
	molecules containing main-group elements,'' {\em Theoretical Chemistry
		Accounts}, vol.~135, Feb. 2016.
	
	\bibitem{pueyo_bellafont_predicting_2017}
	N.~Pueyo~Bellafont, F.~Viñes, W.~Hieringer, and F.~Illas, ``Predicting core
	level binding energies shifts: {Suitability} of the projector augmented wave
	approach as implemented in {VASP},'' {\em Journal of Computational
		Chemistry}, vol.~38, pp.~518--522, Mar. 2017.
	
	\bibitem{takahata_accurate_2010}
	Y.~Takahata, A.~d.~S. Marques, and R.~Custodio, ``Accurate calculation of {C}1s
	core electron binding energies of some carbon hydrates and substituted
	benzenes,'' {\em Journal of Molecular Structure: THEOCHEM}, vol.~959,
	pp.~106--112, Nov. 2010.
	
	\bibitem{oakley_dft/mix:_2018}
	M.~S. Oakley and M.~Klobukowski, ``Δ{DFT}/{MIX}: {A} reliable and efficient
	method for calculating core electron binding energies of large molecules,''
	{\em Journal of Electron Spectroscopy and Related Phenomena}, vol.~227,
	pp.~44--50, Aug. 2018.
	
	\bibitem{cavigliasso_accurate_1999}
	G.~Cavigliasso and D.~P. Chong, ``Accurate density-functional calculation of
	core-electron binding energies by a total-energy difference approach,'' {\em
		The Journal of Chemical Physics}, vol.~111, pp.~9485--9492, Dec. 1999.
	
	\bibitem{pueyo_bellafont_performance_2016}
	N.~Pueyo~Bellafont, F.~Viñes, and F.~Illas, ``Performance of the {TPSS}
	{Functional} on {Predicting} {Core} {Level} {Binding} {Energies} of {Main}
	{Group} {Elements} {Containing} {Molecules}: {A} {Good} {Choice} for
	{Molecules} {Adsorbed} on {Metal} {Surfaces},'' {\em Journal of Chemical
		Theory and Computation}, vol.~12, pp.~324--331, Jan. 2016.
	
	\bibitem{ozaki_absolute_2017}
	T.~Ozaki and C.-C. Lee, ``Absolute {Binding} {Energies} of {Core} {Levels} in
	{Solids} from {First} {Principles},'' {\em Physical Review Letters},
	vol.~118, Jan. 2017.
	
	\bibitem{aoki_accurate_2018}
	T.~Aoki and K.~Ohno, ``Accurate quasiparticle calculation of x-ray
	photoelectron spectra of solids,'' {\em Journal of Physics: Condensed
		Matter}, vol.~30, p.~21LT01, May 2018.
	
	\bibitem{van_setten_assessing_2018}
	M.~J. van Setten, R.~Costa, F.~Viñes, and F.~Illas, ``Assessing \textit{{GW}}
	{Approaches} for {Predicting} {Core} {Level} {Binding} {Energies},'' {\em
		Journal of Chemical Theory and Computation}, vol.~14, pp.~877--883, Feb.
	2018.
	
	\bibitem{golze_core-level_2018}
	D.~Golze, J.~Wilhelm, M.~J. van Setten, and P.~Rinke, ``Core-{Level} {Binding}
	{Energies} from \textit{{GW}} : {An} {Efficient} {Full}-{Frequency}
	{Approach} within a {Localized} {Basis},'' {\em Journal of Chemical Theory
		and Computation}, vol.~14, pp.~4856--4869, Sept. 2018.
	
	\bibitem{sun_strongly_2015}
	J.~Sun, A.~Ruzsinszky, and J.~Perdew, ``Strongly {Constrained} and
	{Appropriately} {Normed} {Semilocal} {Density} {Functional},'' {\em Physical
		Review Letters}, vol.~115, July 2015.
	
	\bibitem{sun_accurate_2016}
	J.~Sun, R.~C. Remsing, Y.~Zhang, Z.~Sun, A.~Ruzsinszky, H.~Peng, Z.~Yang,
	A.~Paul, U.~Waghmare, X.~Wu, M.~L. Klein, and J.~P. Perdew, ``Accurate
	first-principles structures and energies of diversely bonded systems from an
	efficient density functional,'' {\em Nature Chemistry}, vol.~8, pp.~831--836,
	Sept. 2016.
	
	\bibitem{isaacs_performance_2018}
	E.~B. Isaacs and C.~Wolverton, ``Performance of the strongly constrained and
	appropriately normed density functional for solid-state materials,'' {\em
		Physical Review Materials}, vol.~2, June 2018.
	
	\bibitem{marianski_assessing_2016}
	M.~Marianski, A.~Supady, T.~Ingram, M.~Schneider, and C.~Baldauf, ``Assessing
	the {Accuracy} of {Across}-the-{Scale} {Methods} for {Predicting}
	{Carbohydrate} {Conformational} {Energies} for the {Examples} of {Glucose}
	and α-{Maltose},'' {\em Journal of Chemical Theory and Computation},
	vol.~12, pp.~6157--6168, Dec. 2016.
	
	\bibitem{goerigk_look_2017}
	L.~Goerigk, A.~Hansen, C.~Bauer, S.~Ehrlich, A.~Najibi, and S.~Grimme, ``A look
	at the density functional theory zoo with the advanced {GMTKN}55 database for
	general main group thermochemistry, kinetics and noncovalent interactions,''
	{\em Physical Chemistry Chemical Physics}, vol.~19, no.~48, pp.~32184--32215,
	2017.
	
	\bibitem{tozer_molecular_2018}
	D.~J. Tozer and M.~J.~G. Peach, ``Molecular excited states from the {SCAN}
	functional,'' {\em Molecular Physics}, vol.~116, pp.~1504--1511, June 2018.
	
	\bibitem{van_lenthe_relativistic_1994}
	E.~van Lenthe, E.~J. Baerends, and J.~G. Snijders, ``Relativistic total energy
	using regular approximations,'' {\em The Journal of Chemical Physics},
	vol.~101, pp.~9783--9792, Dec. 1994.
	
	\bibitem{faas_zora_1995}
	S.~Faas, J.~Snijders, J.~van Lenthe, E.~van Lenthe, and E.~Baerends, ``The
	{ZORA} formalism applied to the {Dirac}-{Fock} equation,'' {\em Chemical
		Physics Letters}, vol.~246, pp.~632--640, Dec. 1995.
	
	\bibitem{dyall_relativistic_1999}
	K.~G. Dyall and E.~van Lenthe, ``Relativistic regular approximations revisited:
	{An} infinite-order relativistic approximation,'' {\em The Journal of
		Chemical Physics}, vol.~111, pp.~1366--1372, July 1999.
	
	\bibitem{klopper_improved_2000}
	W.~Klopper, J.~H. van Lenthe, and A.~C. Hennum, ``An improved \textit{ab
		initio} relativistic zeroth-order regular approximation correct to order
	1/c2,'' {\em The Journal of Chemical Physics}, vol.~113, pp.~9957--9965, Dec.
	2000.
	
	\bibitem{perdew_density-functional_1982}
	J.~P. Perdew, R.~G. Parr, M.~Levy, and J.~L. Balduz, ``Density-{Functional}
	{Theory} for {Fractional} {Particle} {Number}: {Derivative} {Discontinuities}
	of the {Energy},'' {\em Physical Review Letters}, vol.~49, pp.~1691--1694,
	Dec. 1982.
	
	\bibitem{cohen_insights_2008}
	A.~J. Cohen, P.~Mori-Sanchez, and W.~Yang, ``Insights into {Current}
	{Limitations} of {Density} {Functional} {Theory},'' {\em Science}, vol.~321,
	pp.~792--794, Aug. 2008.
	
	\bibitem{kahk_core_2018}
	J.~M. Kahk and J.~Lischner, ``Core electron binding energies of adsorbates on
	{Cu}(111) from first-principles calculations,'' {\em Physical Chemistry
		Chemical Physics}, vol.~20, no.~48, pp.~30403--30411, 2018.
	
	\bibitem{powell_recommended_2012}
	C.~Powell, ``Recommended {Auger} parameters for 42 elemental solids,'' {\em
		Journal of Electron Spectroscopy and Related Phenomena}, vol.~185, pp.~1--3,
	Mar. 2012.
	
	\bibitem{powell_elemental_1995}
	C.~Powell, ``Elemental binding energies for {X}-ray photoelectron
	spectroscopy,'' {\em Applied Surface Science}, vol.~89, pp.~141--149, June
	1995.
	
	\bibitem{ley_many-body_1975}
	L.~Ley, F.~R. McFeely, S.~P. Kowalczyk, J.~G. Jenkin, and D.~A. Shirley,
	``Many-body effects in x-ray photoemission from magnesium,'' {\em Physical
		Review B}, vol.~11, pp.~600--612, Jan. 1975.
	
	\bibitem{fuggle_xps_1977}
	J.~Fuggle, ``{XPS}, {UPS} {AND} {XAES} studies of oxygen adsorption on
	polycrystalline {Mg} at ∼100 and ∼300 {K},'' {\em Surface Science},
	vol.~69, pp.~581--608, Dec. 1977.
	
	\bibitem{jennison_calculation_1984}
	D.~R. Jennison, P.~Weightman, P.~Hannah, and M.~Davies, ``Calculation of {Mg}
	atom-metals {XPS} and {Auger} shifts using a Δ{SCF} excited atom model,''
	{\em Journal of Physics C: Solid State Physics}, vol.~17, pp.~3701--3710,
	July 1984.
	
	\bibitem{darrah_thomas_valence_1986}
	T.~Darrah~Thomas and P.~Weightman, ``Valence electronic structure of {AuZn} and
	{AuMg} alloys derived from a new way of analyzing {Auger}-parameter shifts,''
	{\em Physical Review B}, vol.~33, pp.~5406--5413, Apr. 1986.
	
	\bibitem{peng_reactions_1988}
	X.~Peng, D.~Edwards, and M.~Barteau, ``Reactions of {O}2 and {H}2o with
	magnesium nitride films,'' {\em Surface Science}, vol.~195, pp.~103--114,
	Jan. 1988.
	
	\bibitem{yoshimura_degradation_2007}
	K.~Yoshimura, Y.~Yamada, S.~Bao, K.~Tajima, and M.~Okada, ``Degradation of
	{Switchable} {Mirror} {Based} on {Mg}–{Ni} {Alloy} {Thin} {Film},'' {\em
		Japanese Journal of Applied Physics}, vol.~46, pp.~4260--4264, July 2007.
	
	\bibitem{fischer_hydrogen_1991}
	A.~Fischer, H.~Köstler, and L.~Schlapbach, ``Hydrogen in magnesium alloys and
	magnesium interfaces: preparation, electronic properties and
	interdiffusion,'' {\em Journal of the Less Common Metals}, vol.~172-174,
	pp.~808--815, Aug. 1991.
	
	\bibitem{roberts_low_2014}
	M.~W. Roberts, ``Low {Energy} {Pathways} and {Precursor} {States} in the
	{Catalytic} {Oxidation} of {Water} and {Carbon} {Dioxide} at {Metal}
	{Surfaces} and {Comparisons} with {Ammonia} {Oxidation},'' {\em Catalysis
		Letters}, vol.~144, pp.~767--776, Mar. 2014.
	
	\bibitem{favaro_subsurface_2017}
	M.~Favaro, H.~Xiao, T.~Cheng, W.~A. Goddard, J.~Yano, and E.~J. Crumlin,
	``Subsurface oxide plays a critical role in {CO} 2 activation by {Cu}(111)
	surfaces to form chemisorbed {CO} 2 , the first step in reduction of {CO}
	2,'' {\em Proceedings of the National Academy of Sciences}, p.~201701405,
	June 2017.
	
	\bibitem{mudiyanselage_importance_2013}
	K.~Mudiyanselage, S.~D. Senanayake, L.~Feria, S.~Kundu, A.~E. Baber,
	J.~Graciani, A.~B. Vidal, S.~Agnoli, J.~Evans, R.~Chang, S.~Axnanda, Z.~Liu,
	J.~F. Sanz, P.~Liu, J.~A. Rodriguez, and D.~J. Stacchiola, ``Importance of
	the {Metal}-{Oxide} {Interface} in {Catalysis}: {In} {Situ} {Studies} of the
	{Water}-{Gas} {Shift} {Reaction} by {Ambient}-{Pressure} {X}-ray
	{Photoelectron} {Spectroscopy},'' {\em Angewandte Chemie International
		Edition}, vol.~52, pp.~5101--5105, Apr. 2013.
	
	\bibitem{eren_catalyst_2015}
	B.~Eren, C.~Heine, H.~Bluhm, G.~A. Somorjai, and M.~Salmeron, ``Catalyst
	{Chemical} {State} during {CO} {Oxidation} {Reaction} on {Cu}(111) {Studied}
	with {Ambient}-{Pressure} {X}-ray {Photoelectron} {Spectroscopy} and {Near}
	{Edge} {X}-ray {Adsorption} {Fine} {Structure} {Spectroscopy},'' {\em Journal
		of the American Chemical Society}, vol.~137, pp.~11186--11190, Aug. 2015.
	
	\bibitem{nakamura_x-ray_1997}
	J.~Nakamura, Y.~Kushida, Y.~Choi, T.~Uchijima, and T.~Fujitani, ``X-ray
	photoelectron spectroscopy and scanning tunnel microscope studies of formate
	species synthesized on {Cu}(111) surfaces,'' {\em Journal of Vacuum Science
		\& Technology A: Vacuum, Surfaces, and Films}, vol.~15, pp.~1568--1571, May
	1997.
	
	\bibitem{yang_fundamental_2010}
	Y.~Yang, J.~Evans, J.~A. Rodriguez, M.~G. White, and P.~Liu, ``Fundamental
	studies of methanol synthesis from {CO}2 hydrogenation on {Cu}(111), {Cu}
	clusters, and {Cu}/{ZnO}(0001̄),'' {\em Physical Chemistry Chemical
		Physics}, vol.~12, no.~33, p.~9909, 2010.
	
	\bibitem{blum_ab_2009}
	V.~Blum, R.~Gehrke, F.~Hanke, P.~Havu, V.~Havu, X.~Ren, K.~Reuter, and
	M.~Scheffler, ``Ab initio molecular simulations with numeric atom-centered
	orbitals,'' {\em Computer Physics Communications}, vol.~180, pp.~2175--2196,
	Nov. 2009.
	
	\bibitem{havu_efficient_2009}
	V.~Havu, V.~Blum, P.~Havu, and M.~Scheffler, ``Efficient integration for
	all-electron electronic structure calculation using numeric basis
	functions,'' {\em Journal of Computational Physics}, vol.~228,
	pp.~8367--8379, Dec. 2009.
	
	\bibitem{strange_automatic_2001}
	R.~Strange, F.~Manby, and P.~Knowles, ``Automatic code generation in density
	functional theory,'' {\em Computer Physics Communications}, vol.~136,
	pp.~310--318, May 2001.
	
	\bibitem{jensen_optimum_2010}
	F.~Jensen, ``The optimum contraction of basis sets for calculating spin–spin
	coupling constants,'' {\em Theoretical Chemistry Accounts}, vol.~126,
	pp.~371--382, Aug. 2010.
	
	\bibitem{huhn_one-hundred-three_2017}
	W.~P. Huhn and V.~Blum, ``One-hundred-three compound band-structure benchmark
	of post-self-consistent spin-orbit coupling treatments in density functional
	theory,'' {\em Physical Review Materials}, vol.~1, p.~033803, Aug. 2017.
	
\end{thebibliography}
\bibliographystyle{ieeetr}

\end{document}